\documentclass[letter]{aa}
\usepackage{txfonts}
\usepackage{graphicx}
\usepackage{natbib}

\usepackage[hidelinks=true]{hyperref}
\usepackage{color}
\definecolor{dark-red}{rgb}{0.4,0.15,0.15}
\definecolor{dark-blue}{rgb}{0.1,0.1,0.6}
\definecolor{medium-blue}{rgb}{0,0,0.5}
\hypersetup{
    colorlinks, linkcolor={dark-red},
    citecolor={dark-blue}, urlcolor={medium-blue}
}

\begin{document}

\title{Formation of a penumbra in a decaying sunspot}

\author{Rohan E. Louis$^{1}$ \and Shibu K. Mathew$^{2}$ \and 
        Klaus G. Puschmann$^{1}$ \and Christian Beck$^{3}$ \and Horst Balthasar$^{1}$}


\institute{Leibniz-Institut f\"ur Astrophysik Potsdam (AIP),
	  An der Sternwarte 16, 14482 Potsdam, Germany \and
          Udaipur Solar Observatory, Physical Research Laboratory,
          Dewali, Badi Road, Udaipur 313004, Rajasthan, India \and
          National Solar Observatory, Sacramento Peak, 
          3010 Coronal Loop, Sunspot, New Mexico 88349, U.S.A.}

\date{Received ... 2012 / Accepted ...}

\abstract
   {Penumbrae are an important characteristic of sunspots, whose formation is intricately related to the nature of sub-photospheric
magnetic fields.}
   {We study the formation of a penumbra in a decaying sunspot and compare its properties with those seen during the development of a 
proto-spot.}
   {High-resolution spectropolarimetric observations of active region NOAA 11283 were obtained from the spectro-polarimeter on board 
{\em Hinode}. These were complemented with full-disk filtergrams of continuum intensity, line-of-sight magnetograms, and 
dopplergrams from the Helioseismic and Magnetic Imager at high cadence.}
   {The formation of a penumbra in the decaying sunspot occurs after the coalescence of the sunspot with a magnetic fragment/pore, 
which initially formed in the quiet Sun close to an emerging flux region. At first, a smaller set of penumbral filaments develop near 
the location of the merger with very bright penumbral grains with intensities of 1.2\,$I_{\textrm{\tiny{QS}}}$, upflows of 4~km~s$^{-1}$, 
and a lifetime of 10~hr. During the decay of these filaments, a larger segment of a penumbra forms at the location of the 
coalescence. These new filaments are characterized by nearly supersonic downflows of 6.5~km~s$^{-1}$ 
that change to a regular Evershed flow nearly 3 hr later.}
    {The coalescence of the pore with the decaying sunspot provided sufficient magnetic flux for the penumbra to form in the sunspot. The 
emerging flux region could have played a decisive role in this process because the formation occurred at the location of the merger and not on the 
opposite side of the sunspot.}

\maketitle

\section{Introduction}
\label{intro}
The radially oriented, filamentary penumbra is a distinct property of sunspots that distinguishes them from pores.
The formation of a penumbra in a sunspot is quite rapid and grows in segments when the magnetic flux of the 
proto-spot is about 1--1.5$\times10^{19}$ Mx \citep{1998ApJ...507..454L}. \citet{2003ApJ...597.1190Y} and 
\citet{2010A&A...512L...1S} observed that individual penumbral filaments form on time scales of 30 min, and that it takes 
nearly 4 hr for the penumbra to cover half the umbral circumference. Transient penumbrae, which do not develop into stable 
penumbral filaments, also form on similar time scales and have been observed in decaying fragments of sunspots \citep{2012ApJ...755...16L}.
Recently, \citet{2012ApJ...747L..18S} reported the existence of an annular zone in the 
chromosphere around a sunspot, which could be a precursor to penumbral formation. As the penumbra develops, it fills this
annular zone, which according to \citet{2012ApJ...747L..18S} is different from the sunspot moat region. 

The formation of the penumbra is immediately followed by the initiation of the Evershed flow (EF) - a nearly horizontal and radial outflow 
of plasma \citep{1909MNRAS..69..454E}. Consequently, the center- and limb-side penumbra are blue- and redshifted, respectively, when the 
sunspot is located off the disk center. However, the early stages of penumbral formation and the late stages of decay are sometimes associated 
with a counter-Evershed flow that has the opposite sign of the EF \citep{2012ASPC..455...61S,2008ApJ...676..698B}. 
While magnetic flux coalesces with the proto-spot at one side, the penumbra was 
seen to form at the {\em opposite} side by \citet{2010AN....331..563S}. The area outside the proto-spot, close to the emerging flux 
region is characterized by elongated granules and tiny magnetic bipoles \citep{2012A&A...537A..19R}. While it is not fully clear 
why the penumbra forms in the manner it does in a developing sunspot, even less is known about the properties of penumbra 
formation during sunspot decay. In this study, we show that mature penumbral filaments can form even during the late stages of 
sunspot decay. In contrast to the findings of \citet{2010A&A...512L...1S}, we observe the formation of a penumbra {\em at} the location of 
coalescence between a decaying sunspot and a tiny pore. 

\begin{figure}[!h]
\vspace{-23pt}
\hspace{2pt}
\centerline{
\includegraphics[angle=90,width = 0.87\columnwidth]{gband-images.eps}
}
\centerline{
\includegraphics[angle=90,width = \columnwidth]{gband-images-closeup.eps}
}
\vspace{-33pt}
\caption{Active region NOAA 11283. Top: G-band images of the AR on September 3 (left panel)
and 4 (right panel). The {\em vertical rectangle} corresponds 
to the field-of-view scanned by the {\em Hinode}/SP. {\bf{LS}}--leading sunspot, {\bf{DS}}--decaying
sunspot, and {\bf{EFR}}--emerging flux region.  The bottom panels correspond to the region 
outlined by the {\em dashed} lines in the top panel. Solar east and north are to the left and to the top.}
\label{figure01}
\end{figure}

\section{Observations}
\label{data}
For this investigation, we employed full-disk solar images from the Helioseismic and Magnetic Imager \citep[HMI;][]
{2012SoPh..275..229S} on board the Solar Dynamics Observatory \citep[SDO;][]{2012SoPh..275....3P} for active 
region NOAA 11283 observed on 2011 September 4. These data include 4k$\times$4k-pixel images of continuum intensity, 
line-of-sight (LOS) magnetograms, and dopplergrams of the photospheric Fe {\sc i} line at 617.3~nm at a cadence of 12 min and 
a spatial sampling of 0\farcs5~pixel$^{-1}$. 

The same active region was also mapped by the {\em Hinode} spectropolarimeter \citep[SP,][] {2001ASPC..236...33L,2008SoPh..249..233I} 
on September 4 from 18:20--19:32 UT. The SP recorded the four Stokes profiles of the Fe {\sc i} lines at 630 nm with a spectral 
sampling of 2.15 pm. An 18$^{\prime\prime}$~$\times$~81$^{\prime\prime}$ field-of-view (FOV) encompasses a part of the 
leading sunspot and a decaying follower sunspot that was repeatedly scanned with a pixel size of 0\farcs16, and an exposure time 
of 4.8 s per slit position (normal map mode). The observations were reduced with the corresponding routines of the 
Solar-Soft package \citep{2013SoPh..tmp....2L}. In 24 hr, the active region traversed from a heliocentric angle
($\Theta$) of 22$^\circ$ to $\Theta=$10$^\circ$ on September 4.

\begin{figure*}[!ht]
\centerline{
\includegraphics[angle=90,width = \textwidth]{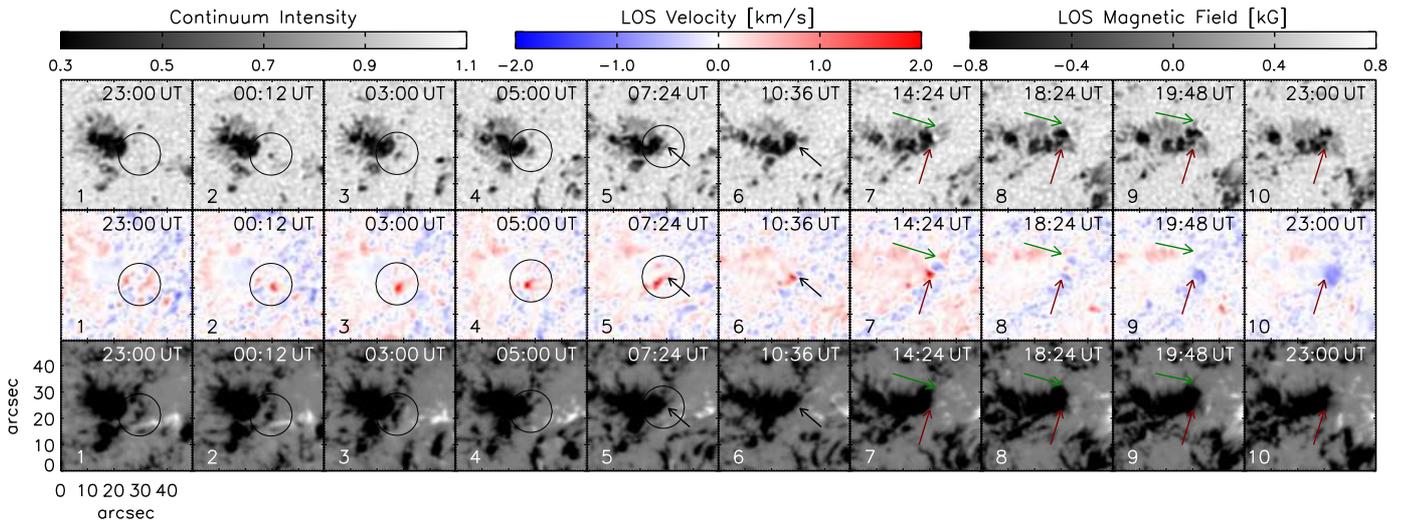}
}
\vspace{-10pt}
\caption{Formation of a penumbra in the decaying sunspot. First, second, and third rows correspond to HMI continuum intensity filtergrams, 
dopplergrams, and LOS magnetograms. All images were scaled according to their respective color bars shown on the 
right. For the dopplergrams, blue and red signify blueshifts and redshifts. The temporal evolution with a larger field of view 
covering a longer time span is shown in the movie available in the online edition. In that movie the square indicates the field of view of 
the panels show in this figure.}
\label{figure02}
\end{figure*}

\section{Results}
\label{result}
\subsection{Penumbra formation from coalescence}
\label{form}
On 2011 September 4, active region NOAA 11283 consisted of a leading sunspot and a decaying follower spot to its 
north. The follower sunspot is devoid of a penumbra on its western half (bottom left panel of Fig.~\ref{figure01}) 
close to the end of September 3. There is significant decay in the follower sunspot, nearly 14~hr later, with 
several light bridges and with the penumbra encircling mostly the northern half of the sunspot. The top 
panels of Fig.~\ref{figure01} also shows an emerging flux region just outside the decaying sunspot to its west.

We followed the evolution of the decaying sunspot using HMI filtergrams of continuum intensity, 
dopplergrams, and LOS magnetograms\footnote{An animation of the HMI data is also available in the online section.} 
as shown in Fig.~\ref{figure02}. The black circle in the top panels indicates an emerging magnetic patch/pore at the beginning of September 4 
(continuum image in panel 2), which is located 7$^{\prime\prime}$ to the west of the decaying sunspot. 
The patch is of negative polarity, the same as the decaying sunspot. This patch shows redshifts (RS) of
$\approx$1.5 km~s$^{-1}$ in the corresponding LOS dopplergram. Panel 4 of Fig.~\ref{figure02} shows the magnetic patch to have 
increased by a factor of four in area nearly 5 hr after emergence and to have a magnetic flux of $-2.8\times10^{19}$~Mx. The patch 
persistently exhibits strong RS of $\approx$2.5~km~s$^{-1}$ and drifts toward the decaying sunspot with 
a speed of 300~m~s$^{-1}$. Nearly 7 hr after emergence, the magnetic patch coalesces with the decaying sunspot
(continuum image in panel 5) while continuing to be strongly redshifted. After merging with the 
decaying sunspot, a part of the magnetic fragment develops into a well-defined penumbral structure (black arrow in panels 
5--6 of Fig.~\ref{figure02}). A close-up view of this penumbral filament is shown in the lower right panel of 
Fig.~\ref{figure01}, indicated by the black circle. The G-band image also shows traces of rudimentary 
penumbrae at the umbra-quiet Sun boundary south of the bright penumbral grain. 
While the bright penumbral grain is 
associated with weak blueshifts (BS) of 470~m~s$^{-1}$, the umbra-penumbra boundary immediately south of it continues to  
exhibit strong RS (dopplergram in panel 6 of Fig.~\ref{figure02}). The green arrow in panels 7--9 of Fig.~\ref{figure02} 
follows the decay of the penumbral filament, and the intensity of the bright penumbral grain exceeds 
that of the quiet Sun (QS) by $\approx$20\%. The lifetime of the transient penumbral filament is $\approx$ 10~hr.

\subsection{Strong redshifts and regular Evershed flow}
\label{ef}
While the penumbral filament described in Sect.~\ref{form} decays, the adjacent region south of it rapidly develops into a distinct 
penumbral section (maroon arrows in panels 7--10 of Fig.~\ref{figure02}) within 4~hr. Prior to its formation, strong RS 
of $\approx$2.5~km~s$^{-1}$ are observed at the umbra-QS boundary (maroon arrow in panel~7). These RS are associated with 
the magnetic fragment that had earlier merged with the decaying sunspot. Once the penumbra has 
matured, we detect a regular EF with BS of about 0.7~km~s$^{-1}$ in the corresponding dopplergrams 
(panels 9--10). Time-lapse movies of the continuum intensity images and dopplergrams reveal that even after the initiation of penumbral 
formation, the strong RS persist for nearly 3 hr before the regular EF starts. The newly formed penumbra shown in 
panel 9 of Fig.~\ref{figure02} contains a magnetic flux of~$\approx-2.0\times10^{19}$~Mx and lies in close proximity to the emerging 
flux region (top right panel of Fig.~\ref{figure01}). 

\subsection{Strong velocities in the penumbra}
\label{velo}
\begin{figure*}[!ht]
\centerline{
\includegraphics[angle=90,width = 0.95\textwidth]{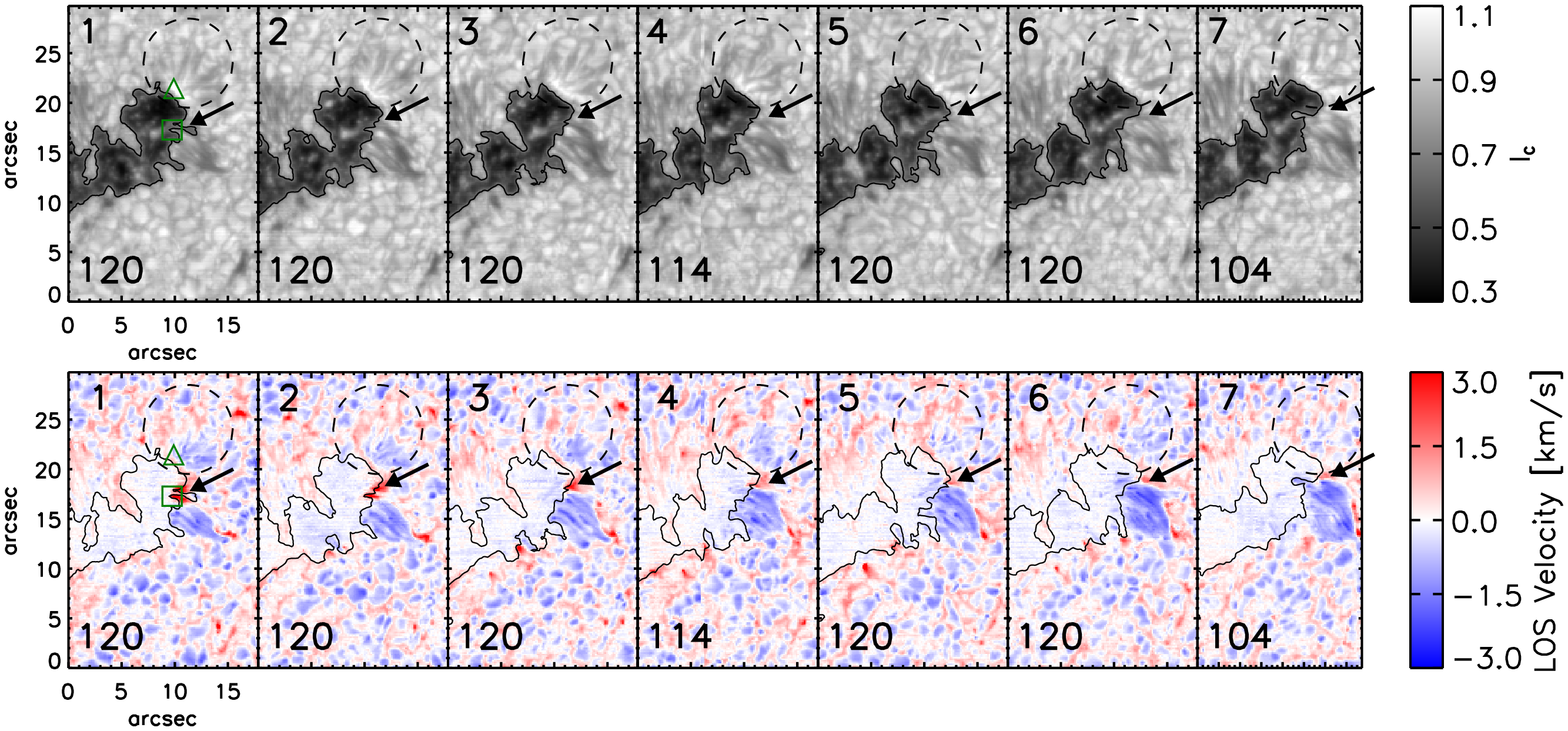}
}
\vspace{-130pt}
\caption{Regular Evershed flow and strong red-shifts in penumbrae from Hinode/SP observations. Top panels -- continuum intensity at 630 nm. 
Bottom panels -- bisector velocity derived at the 70\% intensity level of the 6301.5\AA~ line. Panel numbers are indicated
on the top left corner while the number of scan steps for each panel is shown at the bottom left. The start of the scanning
sequence coincides with panel 8 of Fig.~\ref{figure02}. The {\em green triangle} and {\em square} are centered on pixels whose Stokes
profiles are shown in Fig.~\ref{figure04}.}
\label{figure03}
\vspace{3pt}
\centerline{
\includegraphics[angle=90,width = 0.42\textwidth]{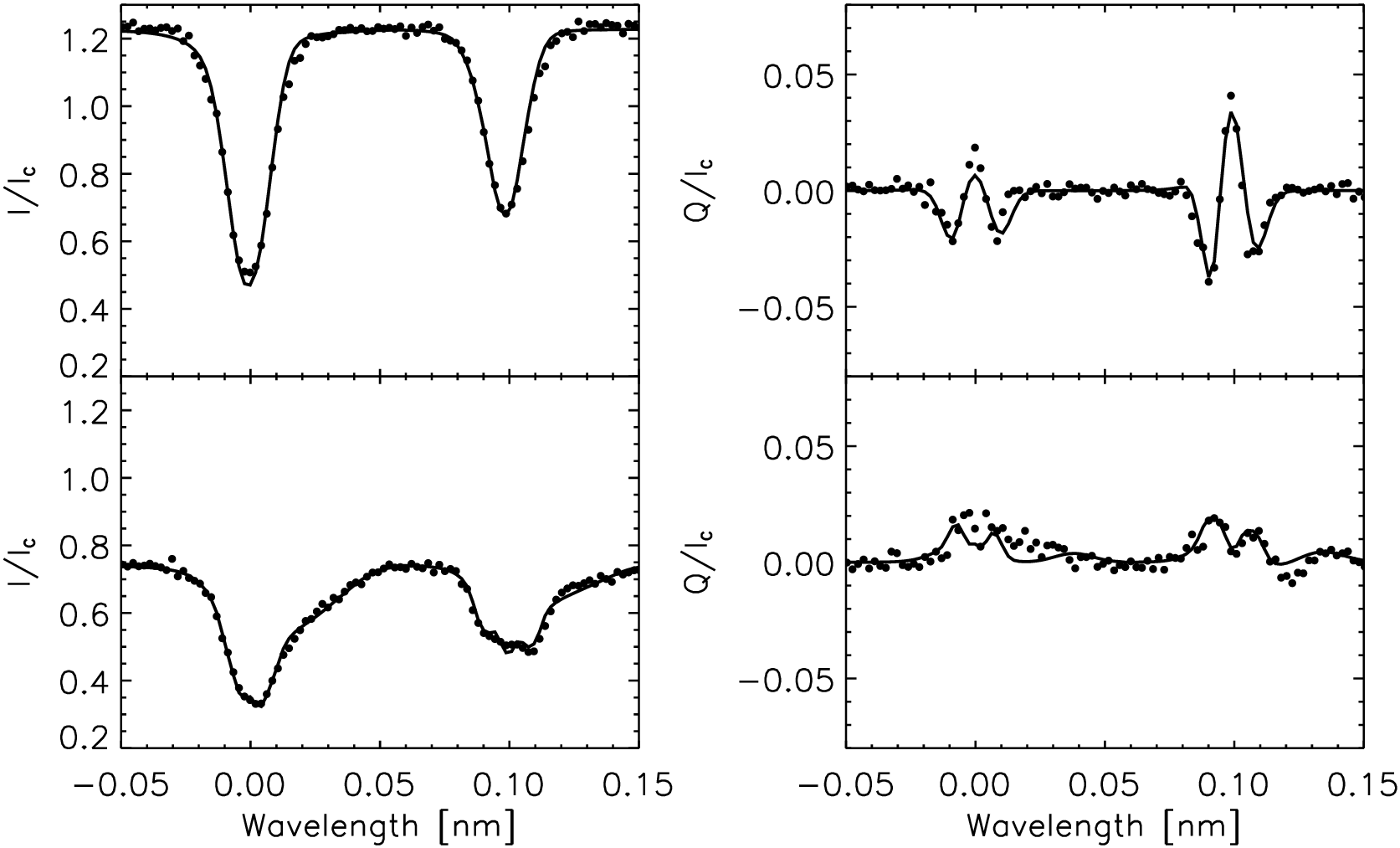}
\hspace{10pt}
\includegraphics[angle=90,width = 0.42\textwidth]{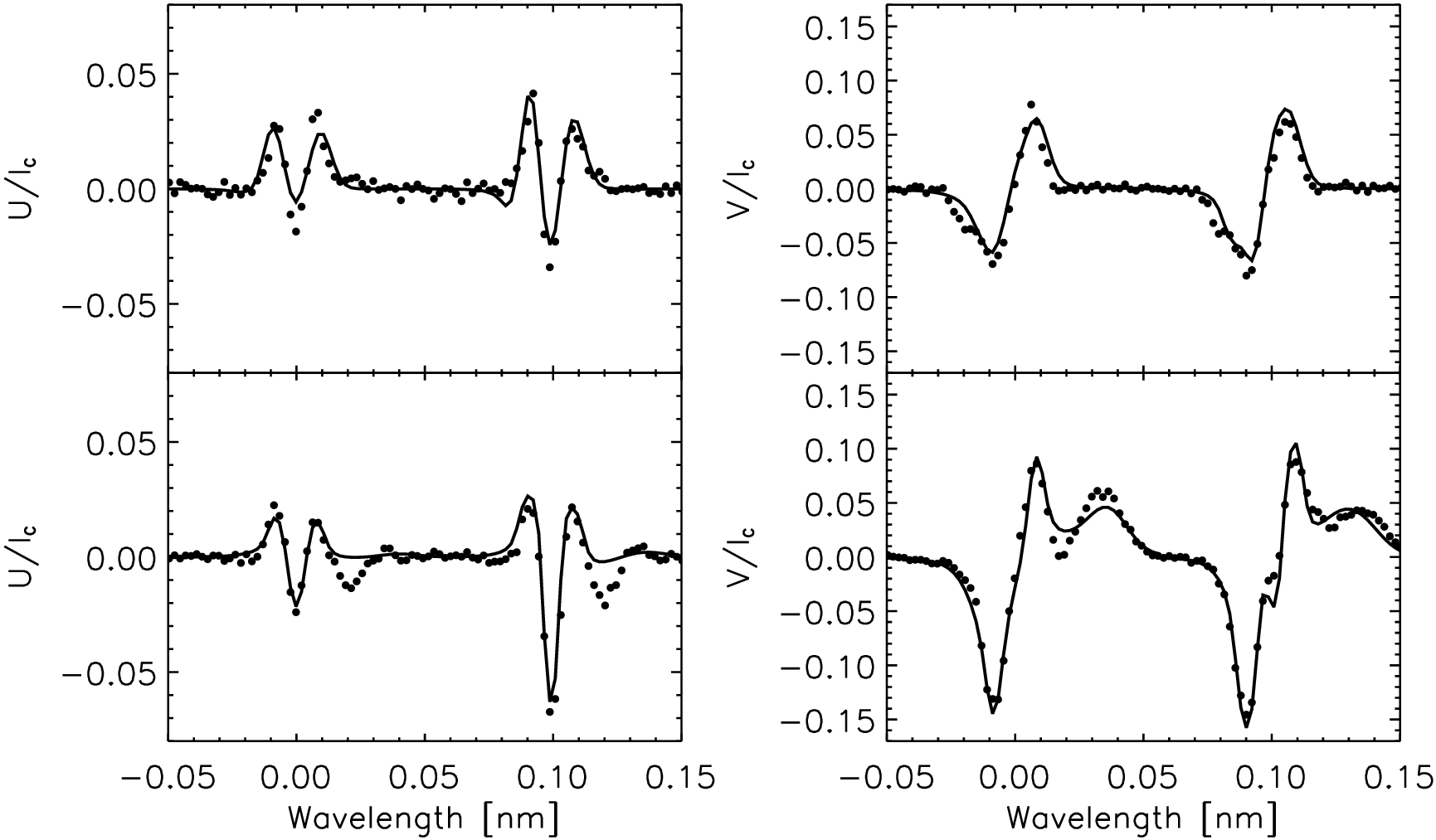}
}
\vspace{-25pt}
\caption{Stokes profiles corresponding to the bright penumbral region (top panel) and the strong downflows 
(bottom panel). The observed and inverted profiles are shown by {\em filled circles} and {\em solid lines}, respectively.}
\label{figure04}
\end{figure*}

Figure~\ref{figure03} shows a sequence of {\em Hinode}/SP scans of the decaying sunspot when it was 
at $\Theta=$14$^\circ$. The bisector LOS-velocities shown in the 
bottom panel were derived at the 70\% intensity level of the Fe 6301.5\AA~line. The figure shows the decay of the penumbral filament 
with the very bright penumbral grain (dashed circle). The continuum images in panels 6--7 clearly show elongated granules that are 
flanked by dark intergranular lanes that are reminiscent of the decaying penumbral structure. These rudimentary filaments are associated 
with BS of up to 1.0~km~s$^{-1}$ (panel 1), which decrease as the structure decays. In comparison, the larger, well-developed 
penumbral section includes both the regular EF and strong RS, as seen in the corresponding LOS velocity maps. 
The EF is associated with BS of nearly 2.0~km~s$^{-1}$ while the RS are about 3.5~km~s$^{-1}$. The black arrow 
points to one location of the RS that lies next to the regular EF. The sequence of the SP scans shows that the 
strong redshifted patch in the penumbra reduces in area, and the velocity decreases to 1.5~km~s$^{-1}$ in about 1 hr.

Figure~\ref{figure04} shows the Stokes profiles (filled circles) emanating from a pixel in the bright penumbral region (green triangle in 
panel 1 of Fig.~\ref{figure03}) and in the redshifted patch (green square in panel 1 of Fig.~\ref{figure03}) in the top and bottom 
panels, respectively. The top panel indicates that the continuum intensity exceeds that of the quiet Sun by $\approx$20\%, which is also seen 
in the HMI images. In addition, the Stokes $V$ profiles of both lines exhibit a small bump in the blue lobe, which is a 
characteristic feature of the photospheric EF. The Stokes $I$  profiles corresponding to the RS (bottom panel of Fig.~\ref{figure04}) 
are characterized by highly inclined red wings, while the $V$ signals show an additional red lobe that signifies a high velocity 
and has the same polarity as the sunspot. The velocities were estimated by inverting the observed profiles with the SIR code 
\citep[Stokes Inversion based on Response functions,][]{1992ApJ...398..375R}. A two-component model atmosphere was assumed with 
height-independent parameters except for the temperature, which was perturbed with two nodes. The observed and inverted profiles are indicated 
by filled circles and solid lines in Fig.~\ref{figure04}. We find that the penumbral filament with the intense brightening at the umbra-penumbra 
boundary has a BS of 3.8~km~s$^{-1}$ in one of the components, whose fill fraction is about 26\,\%. The field strengths and 
inclinations (relative to the solar surface) for the two components are quite similar with values of about 1.1~kG and 117$^\circ$. In 
the redshifted patch, one of the components exhibits near supersonic velocities of about 6.5~km~s$^{-1}$, while the other component 
is at rest. The magnetic field in the redshifted patch is stronger (2.3 kG) and more vertical (135$^\circ$) in both components 
than at the location of the intense brigthening. The additional lobe in the red-wing in Stokes $U$, which is not reproduced by the 
two-component inversion, presumably emanates from a third magnetic component that is nearly horizontal and only weakly ($\sim$0.5~km~s$^{-1}$) 
redshifted.

\section{Summary and conclusions}
\label{summary}
We have analyzed the process of penumbral formation in a decaying sunspot by combining observations from SDO/HMI and the {\em Hinode}/SP. The 
formation of the penumbra is initiated after a magnetic fragment/pore coalesced with the decaying sunspot in the vicinity of an emerging flux 
region. This fragment emerged in the QS about 7$^{\prime\prime}$ to the west of the decaying sunspot. During its motion towards the decaying 
sunspot, the fragment continuously exhibits redshifts exceeding 2~km~s$^{-1}$ that persists even after coalescence. 
Initially, a small set of penumbral filaments forms at the location of the merger and survives for about 10~hr.
During the decay of these filaments, the umbra-penumbra boundary is characterized by blueshifts of nearly 4~km~s$^{-1}$ and 
intensities exceeding those of the QS by 20\%. 

The region south of the decaying penumbral filament rapidly develops into a large penumbral segment with nearly the same 
magnetic flux as the fragment that had earlier merged with the sunspot. These filaments are initially 
associated with nearly supersonic redshifts of 6.5~km~s$^{-1}$, which change after 3 hr to a 
regular EF with relatively weak blueshifts of about 2~km~s$^{-1}$. Since the sunspot was close to disk center at the time of the 
SP observations ($\Theta=14^\circ$), the LOS velocities derived from the bisector method and the inversions practically refer to the vertical 
component of the plasma motion relative to the solar surface. Therefore, blueshifts and redshifts imply upflows and downflows. 
A sonic horizontal radial outflow (inflow) would yield an LOS blueshift (redshift) of only about 1.5~km~s$^{-1}$ at the heliocentric angle of 
the observations.
The downflows and regular EF are seen to coexist in the penumbra. The EF starts as weak upflows in the inner penumbra 
\citep{2006A&A...453.1117B,2008A&A...480..825B,2009A&A...508.1453F,2010ApJ...720.1417P} and ends as downflows in the outer penumbra or beyond the 
sunspot boundary \citep{1997Natur.389...47W,2004A&A...427..319B,2013A&A...550A..97F}. While the downflows associated with the EF have a polarity 
opposite to that of the sunspot, the downflows reported here have the same polarity as the decaying sunspot. It is unclear at this point how these 
strong downflows change to the regular EF and if they have any relation to the supersonic downflows at the umbra-penumbra boundary that are 
possibly related to reconnection effects \citep{2011ApJ...727...49L}.

The formation of penumbrae in a decaying sunspot is similar to and differs from that observed in a developing proto-spot. Penumbrae 
form as a result of an increase in magnetic flux, as in the present case, which was supplied by the magnetic fragment/pore that merged with the 
decaying sunspot. In addition, the penumbra was associated with redshifts during the initial stages of formation, which agrees with 
\citet{2012ASPC..455...61S}. However, contrary to \citet{2010A&A...512L...1S}, we found that the penumbra forms at the location of 
the coalescing fragment, in close proximity to an emerging flux region and not on the opposite side of the sunspot. While the penumbral segment 
was observed until the end of the September 5, the follower sunspot completely decayed by the end of September 6. The emerging flux region developed 
into a nearly complete sunspot 48 hr later. These similarities and differences could be attributed to the emerging flux region close to the 
decaying sunspot, where the growth of penumbrae involves a complex interaction of several individual flux systems in the subphotosphere.
\begin{acknowledgements}
Hinode is a Japanese mission developed and launched by ISAS/JAXA, with NAOJ as domestic partner and 
NASA and STFC (UK) as international partners. It is operated by these agencies in co-operation with 
ESA and NSC (Norway). HMI data are courtesy of NASA/SDO and the HMI science team. They are provided by 
the Joint Science Operations Center -- Science Data Processing at Stanford University. R.E.L is grateful 
for the financial assistance from the German Science Foundation (DFG) under grant DE 787/3-1. R.E.L. 
appreciates the suggestions from P. Venkatakrishnan and Carsten Denker on the manuscript. 
We thank the referee Rolf Schlichenmaier for his valuable comments.
\end{acknowledgements}

\Online

\begin{appendix}
\section{Animation of penumbral formation}
\end{appendix}

\end{document}